\documentclass{nature}
\usepackage{graphicx}

\title{Exploring nine simultaneously occurring transients on April 12th 1950}

\author{Beatriz Villarroel$^{\ast,1,2}$, Geoffrey W. Marcy$^3$, Stefan Geier$^{4,2}$, Alina Streblyanska$^{4,2}$, \\ Enrique Solano Marquez$^{6,7}$, Vitaly N. Andruk$^{8}$, Matthew E. Shultz$^{9}$, Alok C. Gupta$^{10}$, \\ Lars Mattsson$^{1}$}

\begin{document}

\maketitle

\begin{affiliations}
 \item Nordita, KTH Royal Institute of Technology and Stockholm University, Roslagstullsbacken 23, SE-106 91 Stockholm, Sweden
 \item Instituto de Astrof\'isica de Canarias, Avda V\'ia L\'actea S/N, La Laguna, E-38205, Tenerife, Spain
 \item Center for Space Laser Awareness, 3388 Petaluma Hill Rd, Santa Rosa, CA, 95404, USA
 \item Gran Telescopio Canarias (GRANTECAN), Cuesta de San Jos\'{e} s/n, 38712 Bre\~{n}a Baja, La Palma, Spain
 \item Universidad de La Laguna, Departamento de Astrofísica, E-38206 La Laguna, Tenerife, Spain
 \item Departamento de Astrof\'isica, Centro de Astrobiolog\'ia (CSIC/INTA), PO Box 78, E-28691, Villanueva de la Ca\~{n}ada, Spain
 \item Spanish Virtual Observatory
 \item Main Astronomical Observatory of the NAS of Ukraine, 27, Akademika Zabolotnoho St., Kyiv, 03143, Ukraine
 \item Department of Physics and Astronomy, University of Delaware, USA
 \item Aryabhatta Research Institute of Observational Sciences (ARIES), Manora Peak, Nainital, 263 002, India
 
\end{affiliations}

\section*{Abstract}
Nine point sources appeared within half an hour on a region within $\sim$ 10 arcmin of a red-sensitive photographic plate taken in April 1950 as part of the historic Palomar Sky Survey. All nine sources are absent on both previous and later photographic images, and absent in modern surveys with CCD detectors which go several magnitudes deeper. We present deep CCD images with the 10.4-meter Gran Telescopio Canarias (GTC), reaching brightness $r \sim 26$ mag, that reveal possible optical counterparts, although these counterparts could equally well be just chance projections. The incidence of transients in the investigated photographic plate is far higher than expected from known detection rates of optical counterparts to e.g.\ flaring dwarf stars, Fast Radio Bursts (FRBs), Gamma Ray Bursts (GRBs) or microlensing events. One possible explanation is that the plates have been subjected to an unknown type of contamination producing mainly point sources with of varying intensities along with some mechanism of concentration within a radius of $\sim$ 10 arcmin on the plate. If contamination as an explanation can be fully excluded, another possibility is fast (t $<0.5$ s) solar reflections from objects near geosynchronous orbits. An alternative route to confirm the latter scenario is by looking for images from the First Palomar Sky Survey where multiple transients follow a line.

\section*{Introduction}
Modern time-domain surveys exploring the sky with a daily time-resolution have enabled the discovery of many interesting astrophysical transients. Through these surveys, tens of thousands of variable or flaring stars, supernovae, asteroids and variable Active Galactic Nuclei (AGN) have been found, and an increasing number of new sub-classes of transients are continuously expanding the variety of phenomena in the transient sky. Time scales of the transients may vary from events shorter than a few minutes, e.g.\ the optical counterpart (afterglow) of a Gamma Ray Burst (GRB), to those lasting for years, for example the long-term variability of an AGN like e.g. Mrk 590, 3C 454.3 and Mrk 421. Projects like the All-Sky Automated Survey \cite{Pojmanski2002}, Optical Gravitational Lensing Experiment  \cite{Udalski2003} (OGLE) and the Catalina Time Transient Survey \cite{Drake2014} have all contributed to setting limits on the detection rates of transients depending on different wavelengths and depths, using the most modern CCD-based imaging and spectroscopic techniques.

Studying transients on longer time scales requires the use of heterogeneous data sets. Heterogeneous datasets bring their own challenges, as the spectral sensitivity of the filters might differ, or the images may have different depths. The ``Vanishing and Appearing Sources during a Century of Observations'' (VASCO) project\cite{Villarroel2016,Villarroel2020} searches for exotic or rare transients. In the VASCO project, a time baseline of 70 years has been probed using data from the US Naval Observatory Catalogue Catalog \cite{Monet2003} (USNO). The direct advantages are that one rids the data from contamination of artificial satellites (there were no human-made satellites in the early 1950s), and has an opportunity to explore variability on both decadal as well as hourly time scales. Disadvantages include differences in data quality, the presence of different kinds of artifacts, and the difficulty of obtaining responses from many of the original observers from the 1950s regarding questions about potential problems in the observations. 
So far, the project has revealed $\sim 100$ point-like transients \cite{Villarroel2020}, with varying amplitudes, where most have been found on the POSS-I E red plates. Most of them could only be found in one image. Since then, we have further examined the VASCO candidates and find that several images show at least two or three transients.

In this Scientific Report, we report a special case where we find nine objects present at the POSS-I red plate from April 12th 1950 at the Palomar Observatory, which reaches a depth of $r \sim 20$ mag. The plate has plate number XE 325 in the DSS Plate Finder. Whether the effect is a real astrophysical phenomenon or some type of instrumental artifact, the effect is crucial to understand the remaining $\sim$ 100 red transients, as some of them may be caused by similar phenomena. On one of the POSS red plates (XE 325), eight star-like objects and one with possible elongation, appear within a $10 \ast 10$ arcmin field-of-view centered at (RA,DEC) = (212.9291,+26.8311) (eq. J2000.0). We shall be referring to these objects as ``nine simultaneous transients'', despite that we are unsure on whether the object with slight elongation has the same origin as the others. The sources are not seen in the blue exposure 30 minutes earlier, nor in a subsequent red plate (XE 324) obtained six days later (see Figure 1), nor in any subsequent archival images, including the POSS-II red imaging survey.  Even in Pan-STARRS (limiting magnitude $r \sim$ 23.4) and SDSS (limiting magnitude $r \sim$ 23) nothing is seen at the positions of the transients. We investigate possible causes for the nine transients and present new, deep observations of the same field-of-view.

\section*{Method and preliminaries}  
In order to extract the coordinates, FWHMs and $r$ magnitude of each object on the plate, we ran the photometric procedure described by Andruk \& Villarroel et al. (to be submitted). This procedure builds on the methods of Andruk et al. (1995, 2017, 2019) \cite{Andruk1995,Andruk2017,Andruk2019}, and is adapted for POSS-I red emulsions specifically. We also used the procedure to check for other strong transients on the same plate (see Supplementary Information).

In order to gain a better understanding of what we see, we performed observations of the given field in the $r$ and $g$ filters with the GTC 10.4 meter telescope and the OSIRIS instrument on the night from the 25th to the 26th of May 2020. The astrometry was improved with {\it Astrometry.net}. Further, we used the Terapix {\em swarp} procedure to improve the astrometry, using SDSS images as a reference field for zero-point calculations in the astrometrical solutions. Exact details are in the Supp. Info, Sections A.1 and A.2. Out of the 9 sources marked in Fig. 1, {\it only 6 fall into the GTC field of view} -- the two lower right ones fall outside the field of view, as well as the most upper left transient. We WCS-matched the POSS-I and astrometry-corrected GTC images.

\section*{Results}

We manually identify the GTC counterparts to the transients using DS9. For each given transient, we look for counterparts residing within a circle with 5'', 3'' and 1.7'' radius. We perform photometry for these counterparts (Supp. Info, Section A.4). Table 1 summarizes the POSS-I and GTC photometry for the six sources covered by the GTC/OSIRIS observations. The transients and their counterparts are presented in Figures 2 and 3. Proper motion may make it difficult to be sure if one has assigned the right counterpart to the right object. For example, in Figure 2 we see four transients, including the slightly elongated one (nr 5). All four have a counterpart within 5 arcseconds, but only three of these are easily distinguishable (the fourth being fainter). Only one of the counterparts are within a radius of 1.7 arcseconds.

We find the following: within a 5 arcseconds radius, 6 out of 6 transients have at least one counterpart. Within a radius of 3 arcseconds, 3 transients have one or more possible counterparts. Within a radius of 1.7 arcseconds (corresponding to the POSS-1 red images pixel size), 2 transients have counterparts. Given the density of objects within the given GTC field, we need to know whether the probability of finding this many counterparts is significant or not. We used a Sextractor procedure with a threshold of 3 $\sigma$ (corresponding to the visual detectability of the counterparts), discounting objects right at the edge of the plate, to calculate the number of objects (2157) within the field of view of the red image (CCD1), that is $241 \times 515$ arcsec$^{2}$ in size. The average density is thus $\rho=0.0173$ arcsec$^{-2}$. To calculate the probability of finding at least one counterpart within an area of $A$, is $P_1=1-exp(-\rho*A)$.

Hence, the corrected probabilities to have at least one object within 1.7", 3.0", 5.0" are $P_1(1.7")=0.1457$, $P_1(3.0")=0.3876$, $P_1(5.0")=0.744$. For calculating the probability of having at least $m$ counterparts in $n$ transients, with a probability $P_1$, we applied the binomial distribution $P(>=m)=\sum_{i=m}^n [n!/(i!(n-i)!] * P_1^i * (1-P_1)^{n-i}$. In our case, this gives $P(\geq 2)(1.7")=0.170$, $P(\geq 3)(3.0")=0.27$, $P(6)(5.0")=0.169$, which cannot reject the hypothesis of pure chance overlaps. It is not unthinkable, that eventual counterparts to the transients may have moved outside the 5 arcsecond circles during 70 years.

We then examined the colours of the possible counterparts, listed in Table 1. M dwarfs, which are generally very red, usually have significantly redder colours $g - r > 1.5$. The colours of the possible counterparts are oppositely bluer with $g - r < 1$. This excludes the M dwarf hypothesis for the simultaneous transients, and indicates that the counterparts are either star-forming galaxies or F, G or K main sequence stars.

\section*{Discussion}

A number of instrumental issues should be commented upon. The glass cover used during the scanning process can produce many small, false stars. Adding the independently scanned SuperCosmos digitization, we can identify the artifacts in the image (see Supp. Info. A.7.2). Ghost images do not leave a PSF shape on the photographic plate. One may further speculate about the observers having exposed the same plate twice with a partially or completely open shutter. If so, one would expect a high density of extra stars all over the plate.

Non-astronomical effects apart from instrumental errors may arise from elementary particles. We can exclude muons and natural radioactivity, as these particles hit the photographic plates at a steady rate, which disagrees with a few plates having extra stars (had muons and natural radioactivity caused extra stars, the photographic plates would be full of them). We discuss cosmic rays and rare cosmic ray events in detail in Section A.7.5. (Supp. Info.). On the other hand, alpha, beta or gamma particles can certainly leave traces in a photographic plate \cite{Becquerel,Kinoshita,Sisefsky}. An elementary particle entering the emulsion at some incoming angle will expose silver halide crystals within a cylinder that is slanted by that same angle. That slanted cylinder projects to the surface of the plate as an elongated oval shape. Such particles that hit the 103aE emulsion at 30 degrees can, in principle, result in an elongation that is so small (30.7 microns) compared to the FWHM of the seeing PSF that the image might appear nearly symmetric and ``star-like'' (see Supp Info, Section A.7.6).

For example, it has been shown \cite{Webb1949} that the detonation of an atomic bomb in Alamogordo, New Mexico in July 1945 left radioactive contaminants due to the isotope Ce-141, seen as fogged spots, on photographic sensitive x-ray film produced by the Eastman Kodak Company. It was the same company that produced the 103aE and 103aO plates during POSS-I. The radioactive contaminants were transported to the Kodak factory by the wind. Following this incident Kodak developed containers that protected the photographic material from radioactive particles. As Mount Palomar is not too far away from New Mexico, it is not impossible that particles from atomic bomb tests contaminated some of the photographic plates, either while storing them or during observation.

A human source may be to blame. Small local events in an observatory have been known give rise to anomalous detections, e.g. the potassium flare stars \cite{Barbier} that were spectra contaminated by French matches, the discovery of unusual radio bursts, that were caused by microwave ovens \cite{Petroff}. In a recent search for laser signals from Proxima Centauri, it was carefully demonstrated how extraordinary spectral signals reminiscent of lasers can be produced by the presence of a calibration lamp during the observations \cite{Marcy2021}. Humidity can also leave traces in a photographic plate. An observer suffering from hay fever during the spring time in California, might have produced aerosols that expose the photographic plates. But sprayed microdroplets of saliva, peaking around $\sim$ 100 micron in size, do not produce circular shapes on a flat surface, but elongated, wobbly and clustered driplets \cite{Fedorenko}, in sharp contrast with the point sources we see. Pure water driplets can be of sizes ranging from a few tens of microns to almost a millimeter in size, but also have a broader distribution of sizes than what is found in the POSS-1 plates. But maybe, another type of unanticipated contamination or degradation that coincidentally only creates objects with star-like brightness profiles is the guilty party. We investigate the other cases of multiple transients among the $\sim$ 100 VASCO transients and note that some very faint transients can occur in apparently contaminated regions, see Supplementary Information A.7.7. The glass cover during the plate scanning process is a possible source of contamination. For the more luminous transients among the simultaneous transients, the point sources are more difficult to reconcile with the hypothesis of contamination. The best way to exclude the possibility of contamination causing the simultaneous transients is by examining the original photographic plates with a microscope \cite{Greiner1990}. Unfortunately, we have no access to the original POSS-I plates.

We next examine the possibility of naturally occurring transients. The total density of naturally occurring fast transients such as flaring stars, GRBs, or FRBs is estimated as about 1.63 events per square degree per day, up to a limiting magnitude of 23 or 24.7 mag in the $g$ band \cite{Andreoni2019}. The Large Synoptic Survey Telescope expects to find 100,000 short-lived transients per day, which corresponds to about 2.5 events per square degree per day \cite{Bellm2016}. Given that we have 9 transients within 30 minutes inside a $10 \times 10$ arcminute box, no well-known natural transients can explain this phenomenon.

Another possibility are fragmenting asteroids. However, the field is far away from the ecliptic. Could the fragmentation of an asteroid sufficiently far from Earth result in a fragmented shower? Let us, for example, consider an asteroid at the main belt at $\sim$ 1.5 AU from the Earth and 2.5 AU from the Sun. The Earths motion ($\sim$29.9 km/s) dominates over the motion of asteroids. During this time, the fragments must have moved at least $50 \ast 60 \ast 29.9$ kilometers during the 50 minutes
of exposure, resulting in 82 arcseconds long streaks in the image. At a distance of 30 AU (Kuiper belt), the displacement of fragments on the image is around 4 arcsec. At a distance on 0.1 AU from Earth and 1 AU from the Sun, the fragments would have moved roughly 20 arc minutes accross the field. Given the constant illumination of the Sun, long streaks are expected due to the orbital motion. Also, had a fragmentation occured, we'd see a parent body in the blue exposure taken 30 minutes earlier. We see no traces of a parent body in the previous blue exposure, taken half an hour earlier.

But, what if faint and small meteoride fragments burn up right upon atmospheric entry and shortly ``glint’’ as they catch fire? A dust grain as small as 1 gram, can reach visible peak magnitudes around 0 - 2 mag, as it burns up during the course of about 10 seconds. An object entering the Earth’s atmosphere at 100 km can have a wide range of velocities. If the object hits the Earth directly ``face-on’’, it can maximum reach a speed of 72 km/s. Or if the meteoride orbits the Earth, it could instead spiral down at speeds as low as 7.9 km/s. At 100 km above the surface of the Earth, such burning fragments will travel over a large fraction of the sky and extend its trajectory far beyond the observed $10 \ast 10$ arcmin $^{2}$, while reaching elongated shapes from the burning. Assuming a typical velocity of 12 km/s, an object travels 36 000 km during the exposure time, i.e. nearly the circumference of the Earth at $\sim$ 100 km. Such a burning process needs to last well below a fraction of a millisecond to be seen as a point source and otherwise shows up as a elongated streak accross the image. Given the extremely short time scale required to see a glint on the POSS-I images under these circumstances, the expected flux dilution is on the orders of $\sim 10^{7}$, which corresponds to a magnitude dilution of at least 19 magnitudes. Given that the POSS-I plates limiting magnitude is around $\sim$ 20 mag, such an event will not be observed. The time scale is also inconsistent with the observed, longer ablation times for meteorites. We therefore conclude that asteroids, meteorites and similar objects cannot explain the simultaneous transients.

Airplane strobes can flash fast, but they are also accompanied by continuous lights that leave streaks. Nine high-altitude fireflies will doubtfully all flash synchronized in time and motion. However, rotating geosynchronous satellites (or debris) may produce short sub-second glints with the observed amplitudes. In fact, most of the very short-lived transients are satellite glints \cite{Nir2000}. One of the key signatures of small metallic objects with flat, reflective surfaces orbiting our Earth in geosynchronous orbits, is the presence of multiple, very fast and bright glints within the same field-of-view within a few minutes from each other \cite{Nir2000}. Many objects only glint once or twice, other can leave several glints along a line \cite{Schaefer1987}. The only problem with this scenario is that no satellites are known to have existed prior to the Soviet-made Sputnik in 1957, seven years after the appearance of the transients in the 1950 POSS-I image.

Finally, we know that the fact that we have 9 simultaneous transients is inconsistent with any well-known natural explanation. This points towards two possible, but not unique, explanations. The first is high-energy contamination of the photographic plates, most likely by radioactive particles. In an epoch of time when many atomic bomb tests were carried out in the United States and in the Soviet Union, this is not an unreasonable explanation, as the Palomar Observatory in California is not located far from the testing sites in Nevada. However, since no official atomic bomb tests were carried between 1949 and 1951 (which means during the exposure of the plate and during its storage time, since plates were normally developed the same night as the observation), we may also consider an alternative, astronomical explanation.

We first consider the possibility of transients from a high-redshift cluster of galaxies. Recently, the discovery of a possible, but extremely unlikely, GRB in a $z \sim 11$ galaxy \cite{Jiang2020A} with an apparent mag $\sim$ 26 (as faint as our possible counterparts), suggested a high rate of GRBs in the early universe \cite{Jiang2020B}. A cluster of high redshift galaxies cannot explain the 9 simultaneous transients because they should continue to flash after the 1950s, which they do not. A follow-up critique showed that the $z \sim 11$ transient is more likely explained by a Solar system satellite of artificial or natural origin \cite{Steinhardt}. If the transient point sources represent real phenomena in the sky, one must also consider how the simultaneous transients can be so well timed when extended over such a large region of the sky, at least covering $10 \times 10$ arcmin$^2$. In order for the transients to have a some synchronization in turning on or off, the furthest transients must be physically separated by less than 30 light minutes, so that their maximum separation is $d_{s} < 5.4 \ast 10^{11}$ m. The distance $r$ to this phenomenon, can then be calculated as $r < d_{s} / \omega$ , where $\omega$ is expressed in radians. For $\omega = 10'$, this corresponds to $r = 1.9 \times 10^{14}$ meters or 0.02 light years which is inside our Solar System. Small, flat and highly reflective objects at, or near, geosynchronous orbits around the Earth could produce multiple, fast glints. For an object at a geosynchronous orbit moving with a typical speed of 15''/s, the produced glint must be faster than 0.5 s to look like a transient in POSS-I given the average FWHM $\sim$ 7'', meaning the flux dilution of the glint is $\sim$ 9.4 magnitudes during the 50 minutes exposure time. Correcting the magnitudes in Table \ref{PLit} for flux dilution, they reach peak magnitudes $\sim 9 - 11$ mags, which is consistent of what is observed for many reflective pieces that are not larger than a few centimeters in size\cite{Nir2000}. If these are truly glints, we estimate the rate of cases of multiple glints to be $\sim 0.07$ hour$^{-1}$ sky$^{-1}$ i.e. negligible in comparison to the current number of satellite glints an astronomer at a ground-based telescope near the equator sees today, $\sim$ 1800 hour$^{-1}$ sky$^{-1}$ (see Supp. Info., Sec. A.8.1.). Such glints are so rare that they are likely to go unnoticed. As a curiosity, we note that six of the nine transients (including the most central ones), can be explained by two straight lines each holding three transients together. If on top of this already extraordinary scenario, we see the GTC counterparts in an ``off'' state, some of the objects need to follow (artificial) trajectories mimicking stars, in order to compensate for the huge expected proper motion and thus moving at least tens of AU every year. In comparison to this, one may find the first proposed scenario with radioactive bomb particles more down-to-Earth. As this case of simultaneous transients is not unique, contamination cannot be excluded as an explanation.

We acknowledge the interesting surplus of star-like objects in the POSS-I catalogue. It is essential to figure out what is causing them, as several of the general VASCO transients belong to the same category. It is also a showcase of the types of anomalies that may be found in the VASCO citizen science project \cite{Villarroel2020b}. We believe the mystery of the simultaneous transients is a detective story worth of the attention of the astronomical community.

\newpage



\begin{addendum}
 \item Based on observations made with the Gran Telescopio Canarias (GTC), installed in the Spanish Observatorio del Roque de los Muchachos of the Instituto de Astrofísica de Canarias, in the island of La Palma. The authors wish to thank the two anonymous referees that had excellent and constructive suggestions that significantly improved the paper, among them the possibility of a human source of contamination and meteorites. The authors wish to thank Mart\'{i}n L\'{o}pez Corredoira for help with the statistical considerations and Brian McLean for help with retrieving the complete POSS plates in Supp. Info. Table 1. B.V. thanks Martin Ward, Guy Nir, Beatrice Eriksson and Ignacio Trujillo for helpful discussions. This research has made use of the Spanish Virtual Observatory (http://svo.cab.inta-csic.es) supported from the Spanish MINECO/FEDER through grant AyA2017-84089. B.V. is funded by the Swedish Research Council (Vetenskapsr\aa det, grant no. 2017-06372) and is also supported by the The L’Or\'{e}al -
UNESCO For Women in Science Sweden Prize with support of the Young Academy of Sweden. She is also supported by M\"{a}rta och Erik Holmbergs donation.
M.E.S. acknowledges financial support from the Annie Jump Cannon Fellowship, supported by the University of Delaware and endowed by the Mount Cuba Astronomical Observatory. 
 \item[Competing Interests] The authors declare that they have no
competing financial interests.
 \item[Correspondence] Correspondence and requests for materials
should be addressed to B.V.~(email: beatriz.villarroel@su.se).
\end{addendum}

\begin{figure}
   \centering
  \includegraphics[scale=0.2]{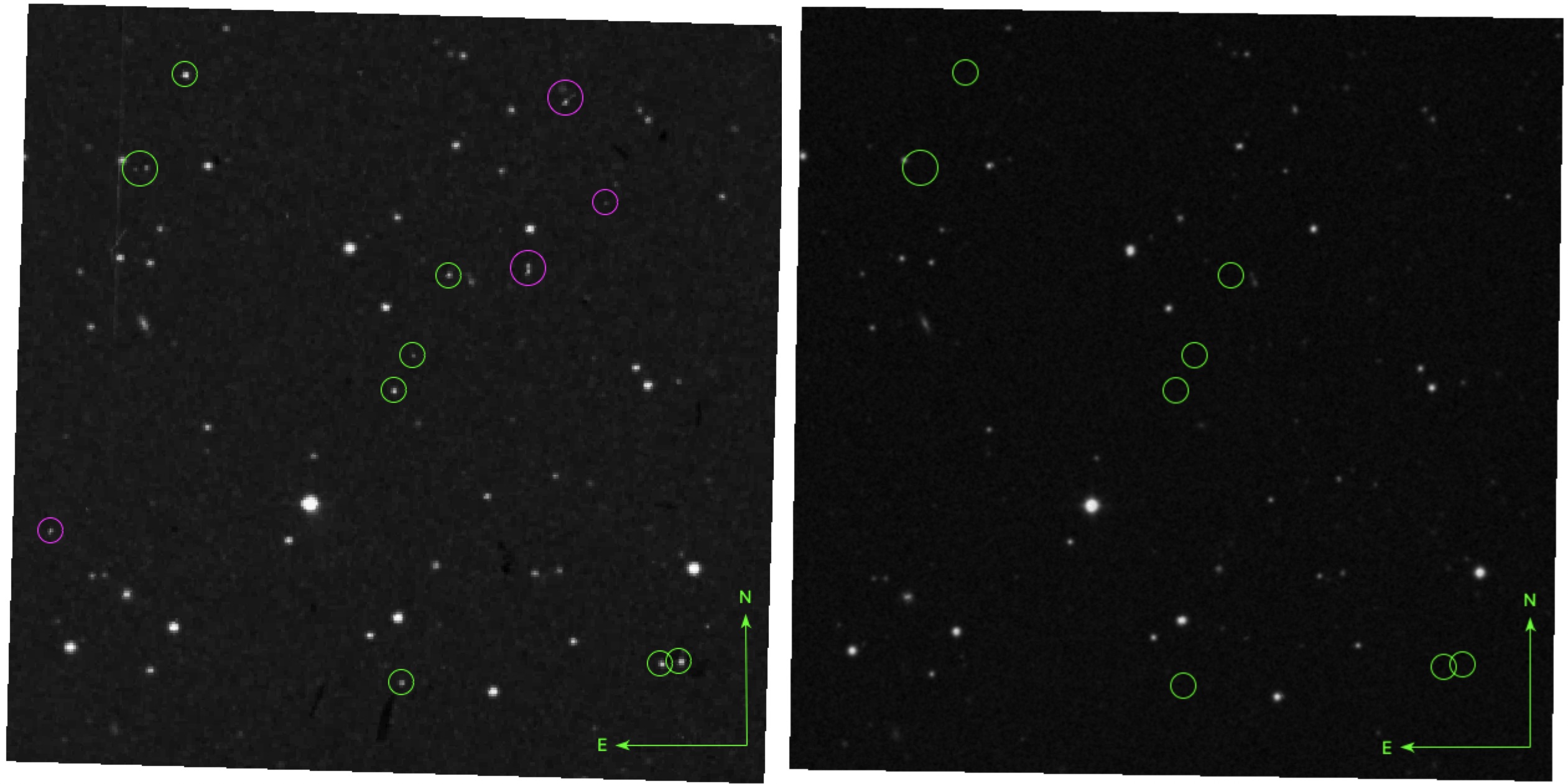}
  \caption{\label{simultaneous} {\bf The same 10 x 10 arcmin field shown in POSS-1 and POSS-2 red bands.} In the POSS-1 image we see a number of objects that cannot be subsequently found, marked with green circles. Purple circles are artifacts during the scanning process. About 9 objects are present in the POSS-I E image (left) from the 12th of April, but not in the POSS-2 image (right) from 1996. One slightly larger circle host two transients. In addition, the 9 objects are neither visible in the blue POSS-1 taken half an hour earlier, nor in a second POSS-1 red image taken six days later on April 18th (see additional images in Supp. Info. Fig.2.) The 9 transients are not caused by a difference in depth or spectral sensitivity. The images are based on the DSS digitizations of the Palomar plates. 
  }
   \end{figure}
   
  \begin{figure*}
  \includegraphics[scale=0.2]{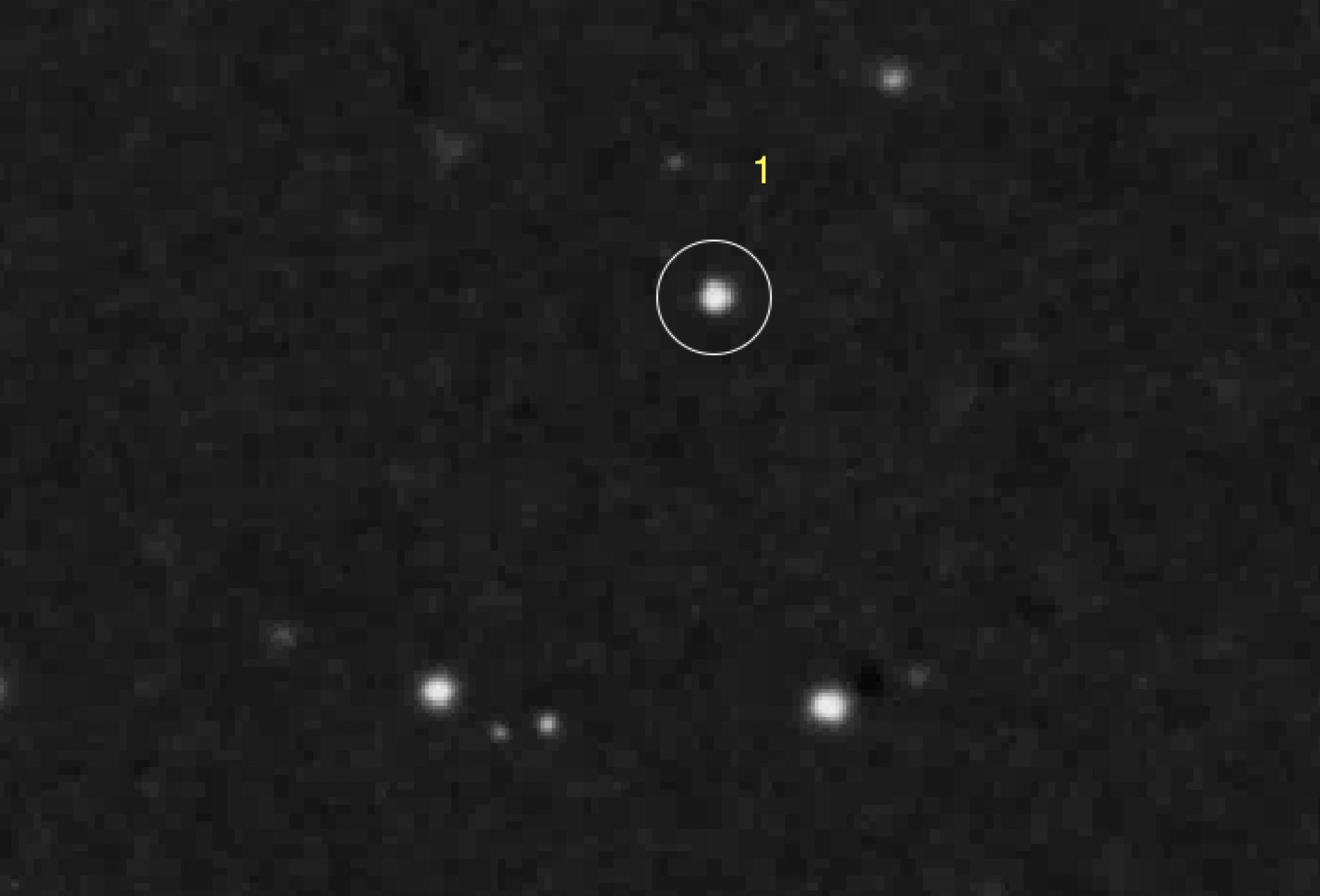}\\
  \includegraphics[scale=0.2]{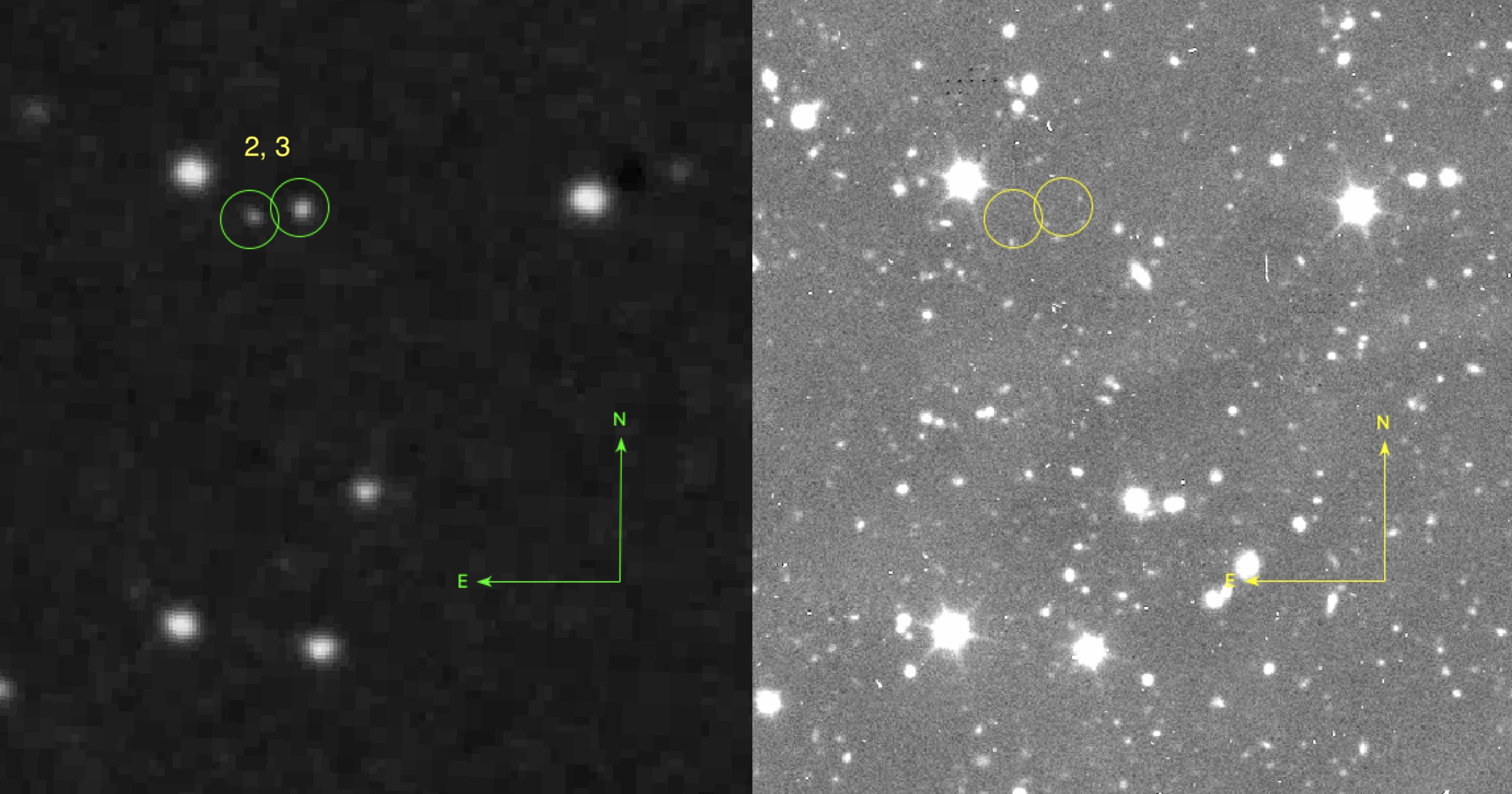}\\
  \includegraphics[scale=0.2]{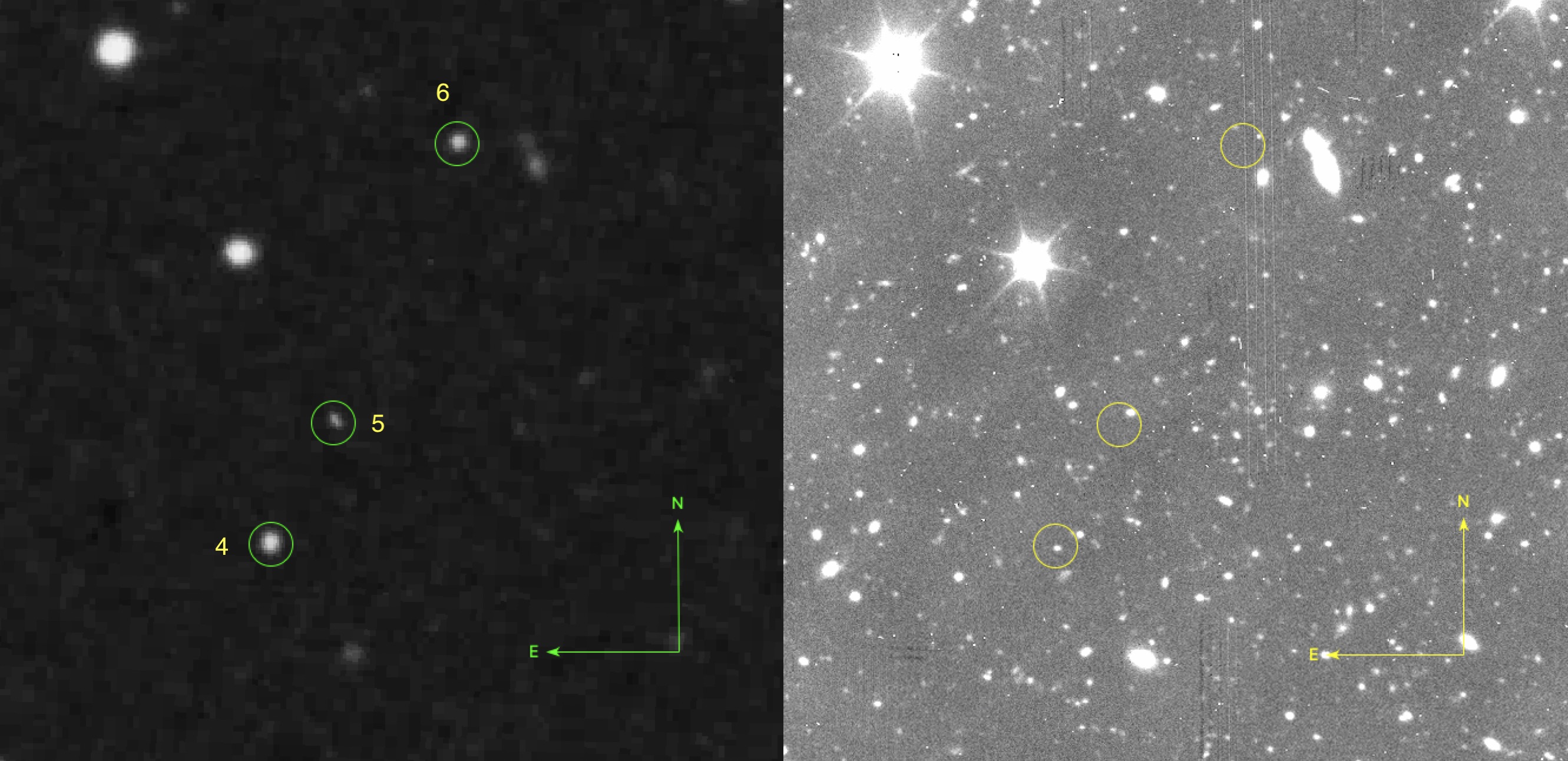}
  \end{figure*}
  \begin{figure*}
     \caption{\label{channel1} {\bf Transients number $1-6$} Upper row: we see transient 1 in POSS-I (marked with a 10 arcsec circle), that is not covered by the GTC observations. Middle and lower row: the left images show the POSS-I red images and the right images show the GTC/OSIRIS g images. The transients are marked with circles with a radius of 5 arcseconds. The number of counterparts is 1 and 3 for transients 2 and 3. The number of counterparts is one each for transients 4, 5 and 6. Transient nr. 5 shows a slight elongation in the POSS-I image. The POSS-I image is based on the high-resolution SuperCosmos digitizations of the Palomar plates.}
   \end{figure*}
   \begin{figure*}
   \includegraphics[scale=0.2]{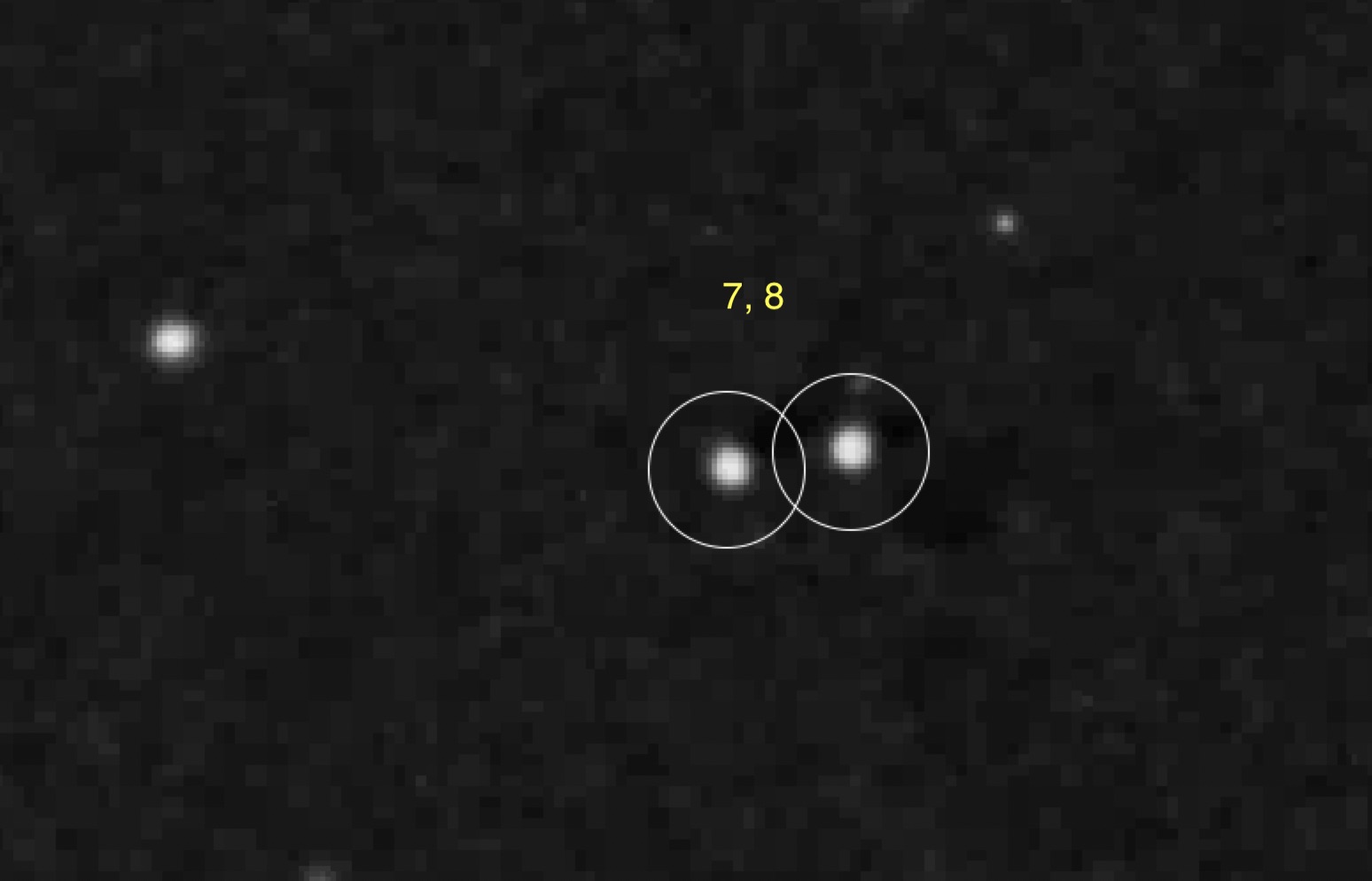}
   \newline
   \includegraphics[scale=0.2]{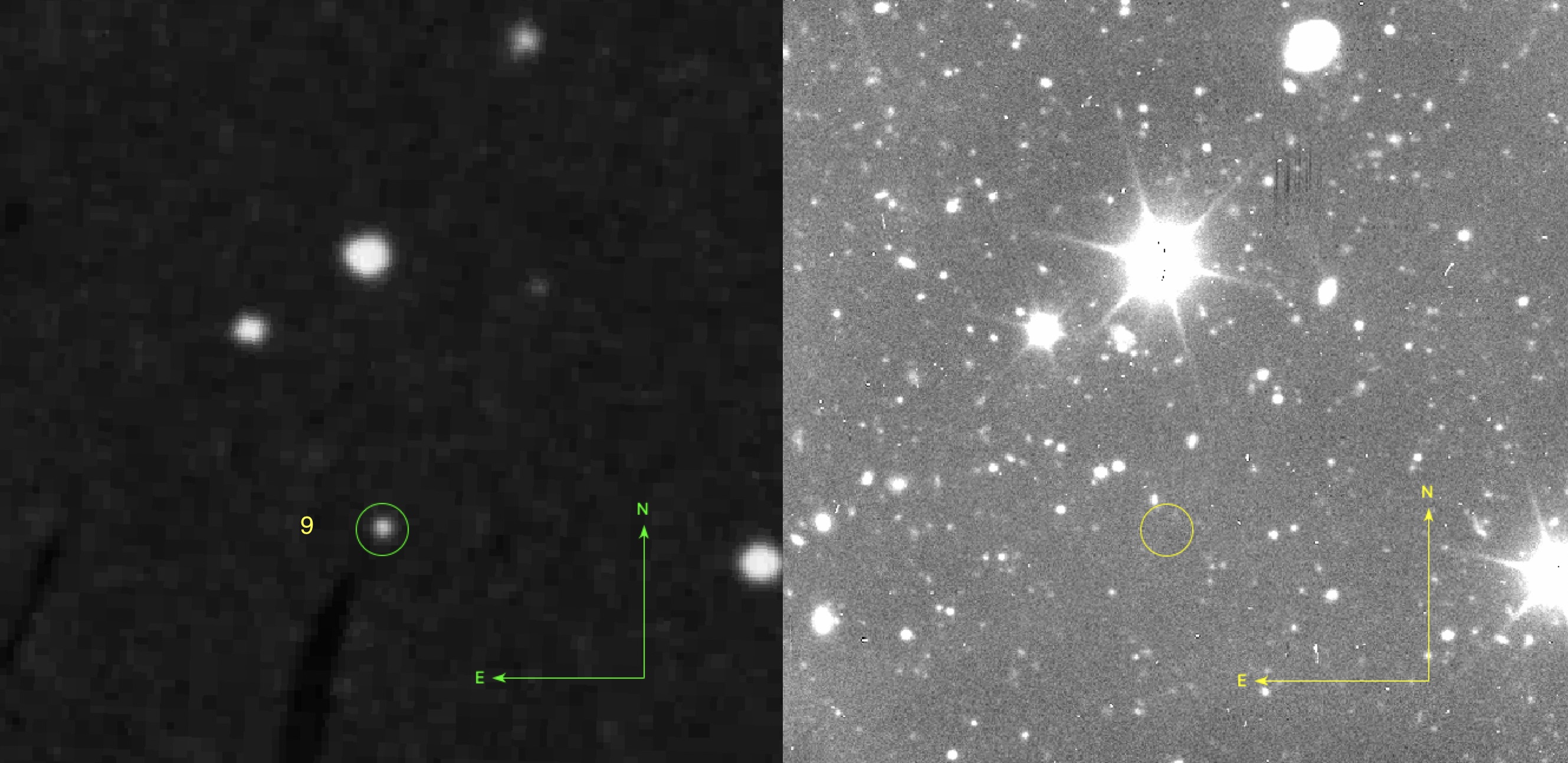}
     \caption{\label{channel2} {\bf Transients number $7-9$.} Upper row: we see transients 7 and 8 in POSS-I (marked with 10 arcsec circles), that are not covered by the GTC observations. Lower row: the left image shows the POSS-I red image and the right image shows the GTC/OSIRIS g image. The transients are marked with circles with a radius of 5 arcseconds. The number of counterparts is 1 for transient 9. The POSS-I are based on the high-resolution SuperCosmos digitizations of the Palomar plates.}
  
   \end{figure*}

   \newpage
   
\begin{table*}[ht]
\caption{Coordinates of simultaneous transients. We display the coordinates and magnitudes of the multiple transients that are covered by the GTC/OSIRIS images, including their counterparts. We see that 6 out of the originally 9 points are covered by the images. When two stars, like object 2 and 3, are too close to each other, the photometry code (that calculates Gaia r magnitudes) fails to detect one of the transients in the POSS-I image, which means we have no photometric information. These are marked with a star (*). Removing outliers with FWHMs larger than 10 pixel in the sample, the average seeing is about 4.8 +/- 0.8 pixel for all objects in the field. Finally, we show the g and r magnitudes (AB) for the closest counterpart.}
\centering
\small
\begin{tabular}{c c c c c c c c}
\hline\hline
\multicolumn{7}{c}{Samples} \\
\hline\hline
number & ra & dec & POSS-I mag & POSS-I FWHM & Counterpart g & Counterpart r & Colour, g-r\\[0.1ex]
2* & 212.9930959* & 26.8802675 & - & - & 26.21 $\pm$ 0.33 & 26.00 $\pm$ 0.29 & 0.21 \\
3 & 212.9902746 & 26.8806478 & 18.80 & 3.876 & 26.31 $\pm$ 0.33 & 26.03 $\pm$ 0.28 & 0.28 \\
4 & 212.9290858 & 26.8318653 & 17.36 & 4.301 & 24.39 $\pm$ 0.21 & 24.11 $\pm$ 0.21 & 0.28 \\
5 & 212.9247558 & 26.8392628 & 19.18 & 3.494 & 23.01 $\pm$ 0.20 & 22.30 $\pm$ 0.17 & 0.71 \\
6 & 212.9157404 & 26.8570081 & 17.99 & 4.237 & 24.93 $\pm$ 0.28 & 24.37 $\pm$ 0.25 & 0.56 \\
9 & 212.9274021 & 26.7673756 & 18.29 & 3.957 & 26.20 $\pm$ 0.34 & 26.89 $\pm$ 0.43 & -0.69 \\[0.1ex]

\hline
\end{tabular}
\label{PLit}
\end{table*}

\end{document}